\RequirePackage{lineno}
\documentclass[twocolumn,letterpaper,aps,prc,superscriptaddress,showpacs,floatfix]{revtex4-1}

\usepackage{graphicx}
\usepackage{dcolumn}
\usepackage{bm}
\usepackage{color}
\usepackage{amsmath}
\usepackage{csquotes}

\begin{document}

\author{R. Belmont and J.L. Nagle}
\affiliation{University of Colorado Boulder, Boulder, Colorado 80309, USA}

\date{\today}

\title{To CME or not to CME? \\ Implications of p+Pb measurements on the chiral magnetic effect in heavy ion collisions}

\begin{abstract}
The Chiral Magnetic Effect (CME) is a fundamental prediction of QCD, and various
observables have been proposed in heavy ion collisions to access this physics.
Recently the CMS Collaboration~\cite{Khachatryan:2016got} has reported results from p+Pb collisions at
5.02 TeV on one such observable, the three-point correlator.  The results are strikingly
similar to those measured at the same particle multiplicity in Pb+Pb collisions,
which have been attributed to the CME.
This similarity, combined with two key assumptions about the magnetic field in p+Pb collisions,
presents a major challenge to the CME picture.
These two assumptions as stated in the CMS paper are (1) that the magnetic field in p+Pb collisions is smaller than
that in Pb+Pb collisions and (2) that the magnetic field direction is uncorrelated with
the flow angle.  We test these two postulates in the Monte Carlo Glauber framework and find
that the magnetic fields are not significantly smaller in central p+Pb collisions,
however the magnetic field direction and the flow angle are indeed uncorrelated.
The second finding alone gives strong evidence
that the three-point correlator signal in Pb+Pb and p+Pb is not an indication of the CME.
Similar measurements in d+Au over a range of energies accessible at RHIC would be elucidating.
In the same calculational framework, we find that even in Pb+Pb collisions, where the magnetic
field direction and the flow angle are correlated, there exist large inhomogeneities that are on the size scale of
topological domains.  These inhomogeneities need to be incorporated in any detailed CME calculation.
\end{abstract}

\pacs{25.75.Gz, 25.75.Gz.Ld}
\maketitle

\section{Introduction}
The search for locally parity (P) violating effects in heavy ion collisions, such as the
Chiral Magnetic Effect (CME), is a major thrust of research in the field of heavy ion
physics; for recent reviews see for example Refs.~\cite{Kharzeev:2015kna,Kharzeev:2015znc}.
The QCD vacuum is highly non-trivial, comprising a spectrum of
topologically distinct states characterized by the Chern-Simons number $N_{CS}$.  When the
gluon field transitions from one topological state to another, a chiral imbalance is
created, described by the relation $\Delta N_{CS} = Q_w = N_L - N_R$, where $Q_w$ is the
topological charge and $N_L$ and $N_R$ are the number of left and right handed quarks,
respectively.  The CME is an electric current either parallel or antiparallel to the magnetic field,
summarized by the relation
\begin{equation}
\vec{J_V} = \frac{N_c}{2\pi}\mu_A\vec{B},
\end{equation}
where $J_V$ is the electric current, $N_c$ is the number of colors (3 in the case of QCD),
$\mu_A$ is the axial chemical
potential (which encodes the strength of the chiral imbalance), and $B$ is the magnetic
field induced by the protons in the target and projectile nuclei.

In semi-central A+A
collisions, the magnetic field is exactly perpendicular to the impact parameter vector in the
case of smooth geometry.  Fluctuations in the nucleon positions are expected to cause small
fluctuations in this relation.  The conventional Fourier series~\cite{Voloshin:1994mz} used to
describe the azimuthal distribution of particles explicitly omits the P-odd sine terms,
but it is straightforward~\cite{Voloshin:2004vk} to add them.  In this case one has
\begin{equation}
\frac{dN}{d\phi} \propto 1 + 2\sum_{n=1}^{\infty} [v_n \cos(n(\phi-\psi)) + a_n\sin(n(\phi-\psi))],
\end{equation}
where $\psi$ is some symmetry plane
(different symmetry planes can be chosen to address different physical mechanisms),
and $v_n = \langle \cos(n(\phi-\psi))\rangle$ are the familiar harmonic coefficients and
$a_n = \langle \sin(n(\phi-\psi))\rangle$ are the P-odd terms.  In the case of the CME,
the first term $a_1$ is the term of interest.  For a given topological charge, the positive
and negative particles travel in opposite directions and therefore $a_1^+$ and $a_1^-$
should have opposite sign.  However, the sign of either of them depends on the sign of the
topological charge, which fluctuates event-by-event about an average of
$\langle Q_w\rangle~=~0$.  To that end, Voloshin proposed~\cite{Voloshin:2004vk} to study
2-particle correlations, where one always expects $\langle a_1^{\pm}a_1^{\pm}\rangle~>~0$
and $\langle a_1^{\pm}a_1^{\mp}\rangle~<~0$.  To measure these quantities, one can measure
the three-point correlator
\begin{equation}
\langle \cos(\phi_{\alpha}+\phi_{\beta}-2\psi_{RP}) \rangle =
\langle v_1^{\alpha} v_1^{\beta} \rangle - \langle a_1^{\alpha} a_1^{\beta}\rangle,
\end{equation}
where $\phi_{\alpha}$ is the azimuthal angle of one particle,
$\phi_{\beta}$ is the azimuthal angle of a second particle,
and $\psi_{RP}$ is the reaction plane angle, which is defined as pointing
along the direction of the impact parameter vector.
The correlator is a three-point correlator in that the first two points are the two particles of
interest and the third point is the symmetry plane.  In cases where one uses a single particle
to determine the symmetry plane, one has the more specific case of a three-particle correlator.
Note that for clarity we omit non-flow and assume the magnetic field direction is perfectly correlated with
the reaction plane angle $\psi_{RP}$ in this equation.
In experiment, one can neither measure the magnetic field direction nor the reaction plane angle
directly.  Instead we measure the second harmonic flow plane, and use that in the three-point
correlator as
\begin{equation}
\langle \cos(\phi_{\alpha}+\phi_{\beta}-2\psi_{2}) \rangle,
\end{equation}
where $\psi_2$ is the second harmonic event plane.  In nuclear collisions,
$\psi_2$ is strongly correlated with $\psi_{RP}$, but they are not equal due
to event by event fluctuations in the nucleon positions~\cite{Alver:2006wh}.
Assuming the measured $\psi_2$ is correlated with the magnetic field direction,
as expected in Pb+Pb collisions, one will get a net contribution to the correlator.
In contrast, background correlations that are independent of the magnetic field and
$\psi_2$ should cancel out because the CME correlator is a difference between
in- and out-of-plane quantities.

The first measurement of the CME correlator was performed by the STAR Collaboration~\cite{Abelev:2009ad} in Au+Au
collisions at $\sqrt{s_{NN}}$ = 200 GeV, and revealed a non-zero signal.
Following that, the ALICE Collaboration made a comparable measurement
in Pb+Pb collisions at $\sqrt{s_{NN}}$ = 2.76 TeV~\cite{Abelev:2012pa}.
The two results are striking in
that the correlator strength was almost identical at the two energies.  Based on simple expectations from the life time of the magnetic field, which
is much shorter at the LHC, the correlation strength was expected to be smaller.
Conversely, the radial and anisotropic flow are comparable between the two, suggesting
that if any CME were present, it could be subdominant to background correlations.  However,
it has also been observed that, in the presence of an electrically conducting
medium, the field lifetime could be extended considerably~\cite{McLerran:2013hla}.  This
could plausibly allow the correlation strength from the CME to be similar at
different heavy ion collision energies.

More recently, it has been observed by the STAR Collaboration~\cite{Adamczyk:2016eux} that the charge dependence of
$v_1$ measured in asymmetric Cu+Au collisions, where there is a strong electric field
pointing in the direction of the impact parameter, indicates that the percentage of quarks
present at early times while the fields are still strong is only about 10\%, which places
constraints on the CME.

Very recently, the CMS Collaboration~\cite{Khachatryan:2016got} has presented the first measurement of the CME correlator
in p+Pb collisions at $\sqrt{s_{NN}}$ = 5.02 TeV, and shown them to be remarkably consistent with the
results in Pb+Pb at $\sqrt{s_{NN}}$ = 5.02 TeV at the same particle multiplicity.  Given that naively one expects no CME in p+Pb collisions,
this places the strongest constraints yet on the level of CME that can contribute to the
observed correlations.  However, the argument they present that one expects no CME in p+Pb
hinges on two postulates: firstly, that the magnetic field in p+Pb collisions is smaller than
that in Pb+Pb; and secondly that the magnetic field is uncorrelated with the flow plane angle $\psi_2$.
In this paper we test these two postulates.

\section{Monte Carlo Glauber and magnetic field calculation}
A detailed calculation of the spatial and temporal dependence of the magnetic field
present in heavy ion collisions and its interaction with the quarks when present can be
complicated.  However, in this case we want to compare the relative magnetic
field magnitude at the initial collision time in the participant region and its
orientation between Pb+Pb and p+Pb interactions.  For this purpose, we utilize a modified
version of the Monte Carlo Glauber code as detailed in Ref.~\cite{Loizides:2014vua}.

We use the standard Woods-Saxon parameters for the Pb nucleus in distributing the
nucleons, including a hard-core repulsive interaction (i.e. nucleons are required to be at
least $d>$~0.4~fm apart based on their center positions).  We assume that the distribution
of protons and neutrons are given by the same function (i.e. no neutron skin is
considered).  For each event, we utilize a nucleon-nucleon inelastic cross section of
60 millibarns and determine if nucleons interact in the black disk picture.  For each
Pb+Pb or p+Pb event, we determine the center-of-mass of the participating nucleons,
defined as nucleons with at least one interaction.  We then calculate the magnitude
($\varepsilon_2$) and angle ($\psi_2$) of the eccentricity by
\begin{eqnarray}
\varepsilon_{2} &=& \frac{\sqrt{\left<r^{2}\cos(2\phi)\right>^{2} + \left<r^{2}\sin(2\phi)\right>^{2}}}{\left<r^{2}\right>}, \label{eq:epsilon2} \\
\psi_{2} &=& \frac{\mathrm{atan2}\left(\left<r^{2}\sin(2\phi)\right>,\left<r^{2}\cos(2\phi)\right>\right)}{2} + \frac{\pi}{2}, \label{eq:psi2}
\end{eqnarray}
where $r$ is the displacement of the participating nucleon from the center-of-mass,
$\phi$ is the angle of the participating nucleon in the transverse plane, and
the brackets indicate an average over all participants.

In this manuscript we calculate only the peak field strength, i.e. the field strength at
the time of impact.  In Ref.~\cite{Voronyuk:2011jd}, they use a dynamical model to
calculate the time dependence of the field strength.  The field strength they determine at
$t=0$ is consistent with our calculations.  Relatedly, the same model for the time
evolution of the fields can be used to evaluate possible background contributions to the
three-point correlator~\cite{Toneev:2012zx}.

\begin{figure*}[ht!]
\centering
\includegraphics[width=0.49\linewidth]{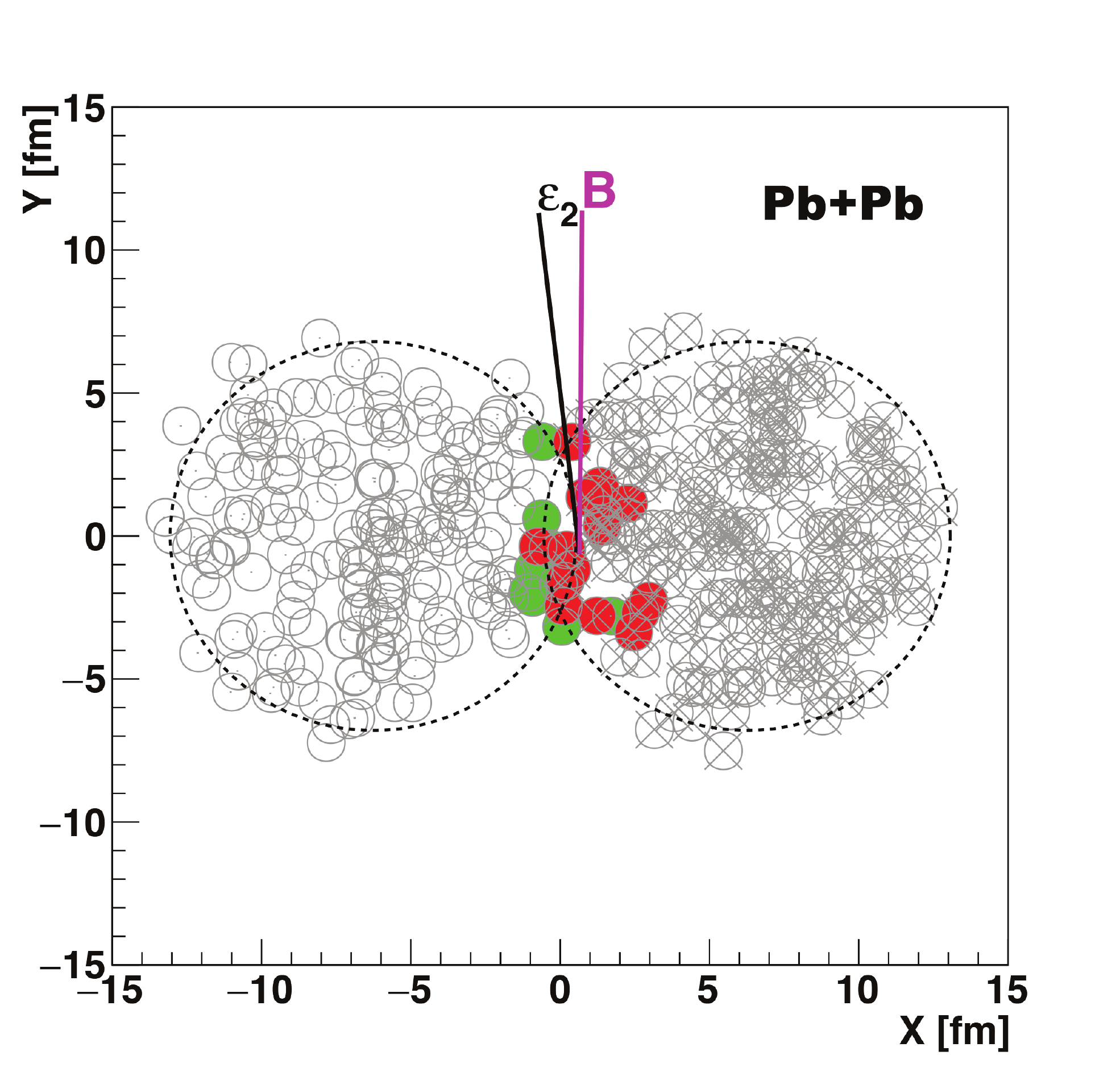}
\includegraphics[width=0.49\linewidth]{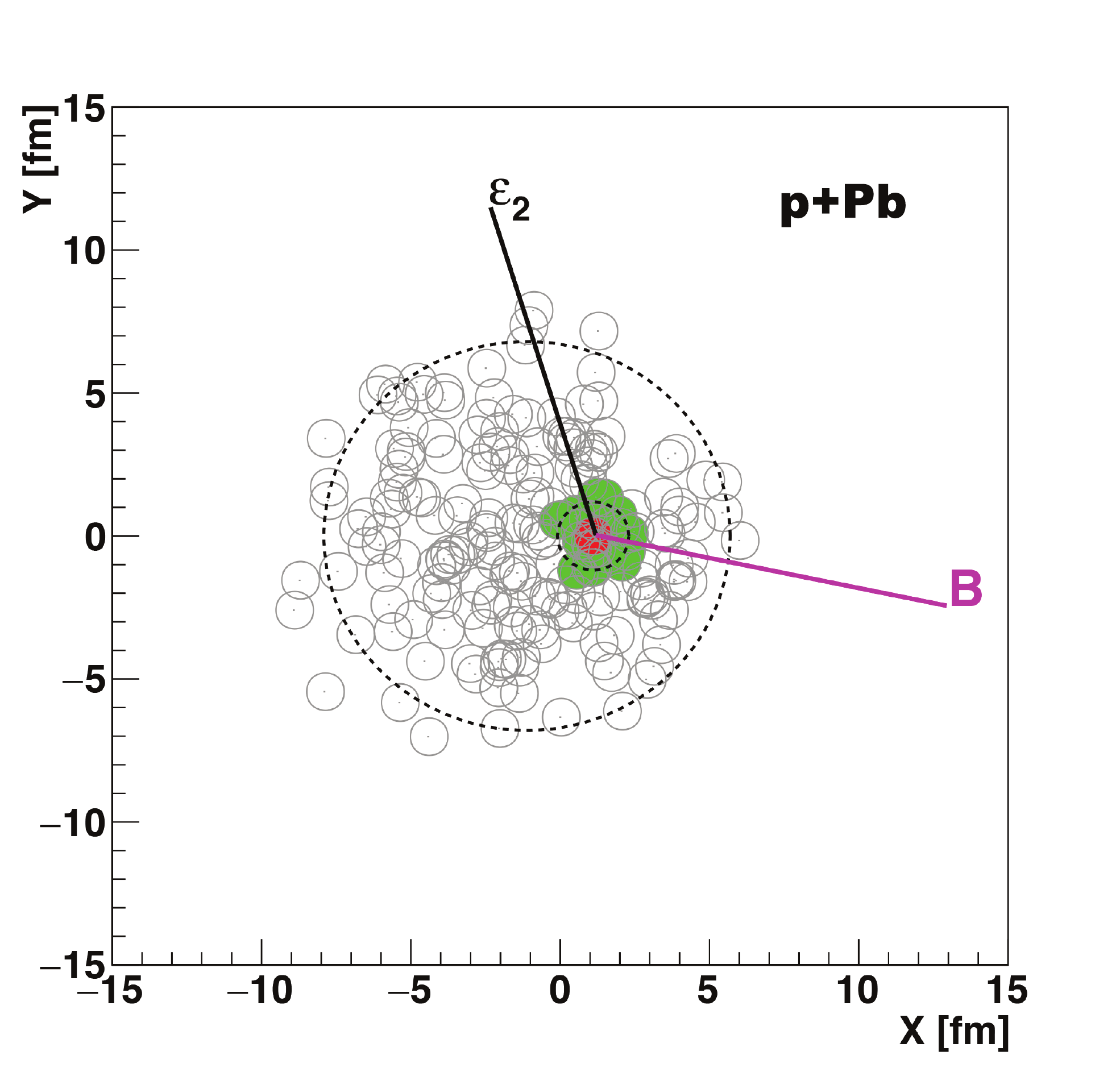}
\caption{
  Single event display from an Monte Carlo Glauber event of a
  peripheral Pb+Pb (left panel) and a central
  p+Pb (right panel) collision at 5.02 TeV.
  The open gray (filled green) circles indicate spectator nucleons (participating protons)
  traveling in the positive $z$-direction, and the open gray (filled red) circles with
  crosses indicate spectator nucleons (participating protons) traveling in the negative
  $z$-direction.
  In each panel, the calculated magnetic field vector is shown as a solid magenta line and the long axis of the
  participant eccentricity is shown as a solid black line.}
\label{fig:event_display}
\end{figure*}

One example Pb+Pb peripheral event is shown in the left panel of Figure~\ref{fig:event_display} in the transverse
plane.  The gray open circles are the nucleon spectators from both nuclei.  The left
nucleus is moving out of the page, as indicated by the dots in the center of the circles.
The right nucleus is moving into the page, as indicated by the crosses in the center of
the circles.  The green (red) filled circles indicate participating nucleons moving out
(into) the page.  The black arrow indicates the orientation of the eccentricity along the
long axis and is drawn from the center-of-mass of the participating nucleons.

The magnetic field is calculated specifically at the center-of-mass position of the
participating nucleons.  We assume a Gaussian spatial distribution with $\sigma$ = 0.4 fm for
the distribution of the electric charge for each proton.  We note that following the
procedure in Ref.~\cite{Bzdak:2011yy}, we checked the calculation results using a point
charge at the center of each proton and excluding protons with $r<$ 0.3 fm of the point
where the field is determined, and found similar results in all cases.
To determine the magnetic field, we use the Biot-Savart law for moving point charges:
\begin{eqnarray}
\vec{E}&=&\frac{q}{4\pi\varepsilon_{0}}\frac{1-v^{2}/c^{2}}{(1-v^{2}\sin^{2}\theta/c^{2})^{3/2}} \frac{\hat{r}'}{\vec{r}^{2}}, \\
\vec{B}&=&\frac{1}{c^{2}}\vec{v} \times \vec{E},
\end{eqnarray}
where $\vec{E}$ is the electric field, $\vec{B}$ is the magnetic field, $q$ is the charge
of the proton, $\varepsilon_0$ is the electric permitivity of free space, $v$ is the velocity
of the proton, $c$ is the speed of light, $\vec{r}$ is the displacement between the proton
and the center-of-mass of the participating nucleons, and $\theta$ is the angle between
the velocity vector and the displacement vector, which is exactly 90$^{\circ}$ at the
moment of impact of the two colliding nuclei.
The vector direction of the magnetic field is shown in the example Pb+Pb interaction in
the left panel of Figure~\ref{fig:event_display}.

In this particular event, the magnetic field is oriented upwards, which is the expectation
in the absence of fluctuations in the positions of the protons.  It is also true in this
one event that the long axis of the eccentricity is aligned closely with the magnetic
field.  Thus, for this event, there is a significant magnetic field along this long axis
and a very small magnetic field perpendicular to it.  This is the type of configuration that makes
the CME maximally observable with the three-point correlator.

We show in the right panel of Figure~\ref{fig:event_display} an example p+Pb interaction
where we again calculate the long axis of the eccentricity and magnetic field vector in
the identical framework.  In this example interaction, the magnetic field and eccentricity
long axis are almost perpendicular.  In
addition, the magnetic field vector itself, due to fluctuations in the positions of the
protons (particularly those closest to the participant center-of-mass), is not along the
expected direction (i.e. expected for the case of a smooth charge distributed nucleus).

\section{Results}
\begin{figure*}[ht!]
\centering
\includegraphics[width=0.95\linewidth]{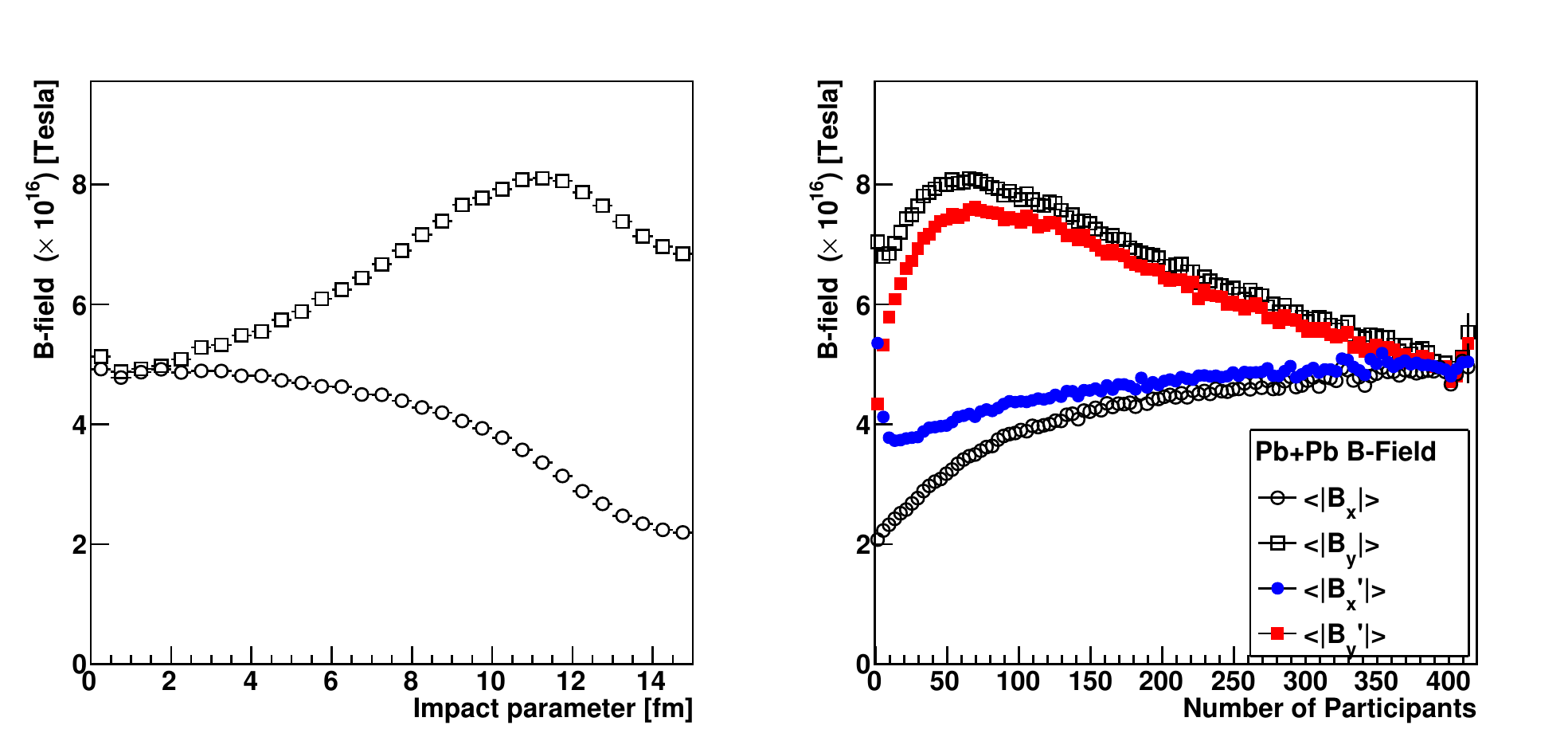}
\caption{Magnetic field components parallel ($|B_x|$, indicated by black open circles) and
  perpendicular ($|B_y|$, black open squares) to the impact parameter vector as a function of
  impact parameter (left panel) and $N_{part}$ (right panel) in Pb+Pb collisions.  The
  filled red squares on the right panel indicate the rotated components parallel to the eccentricity direction ($|B_x'|$) and
  the filled blue circles rotated component perpendicular to the eccentricity direction ($|B_y'|$).}
\label{fig:bfield_pbpb}
\end{figure*}

In order to fully quantify these effects, we sample over one million Pb+Pb and one million
p+Pb Monte Carlo Glauber events.  First, we discuss the Pb+Pb results.  In Figure~\ref{fig:bfield_pbpb},
the left panel shows the mean of the absolute value of the magnetic field oriented along
the $x$-axis $\left< |B_{x}| \right>$ (open circles) and the $y$-axis
$\left<|B_{y}|\right>$ (open squares) as a function of the Pb+Pb collision impact
parameter.  Note that the impact parameter vector is always along the $x$-axis.  The
magnetic field is shown in units of Tesla.
Commonly in the literature the quantity $\hbar e B/c^2$ is reported,
which gives an equivalence $10^{15}~\mathrm{T}~\leftrightarrow~3.0366~m_{\pi}^2$,
where $m_{\pi}$ is the mass of the charged pion (139.57~MeV/$c^2$).

As expected, in peripheral Pb+Pb events, there is a large mean magnetic field oriented in
the $y$-direction, and a rather small mean magnetic field oriented in the $x$-direction.  Note that if
we did not calculate the mean of the absolute value, the mean magnetic field in the
$x$-direction would be zero with as many events fluctuating to have a positive and negative
field along this axis.  In the most central ($b$ close to zero) events, the two magnetic
field components have equal mean values since the magnetic field is entirely due to
fluctuations in the proton positions.  In the right panel, we show the same result, now as
a function of the number of participating nucleons, which is related to the number of
particles produced in the event.

In addition, in the right panel of Figure~\ref{fig:bfield_pbpb}, we show the magnetic field mean values
now oriented along the long axis of the eccentricity (shown as a black arrow labeled
$\varepsilon_{2}$ in Figure~\ref{fig:event_display}), referred to as $\left< |B_{y}'| \right>$, and in
the perpendicular direction, referred to as $\left< |B_{x}'| \right>$.  It is striking
that due to significant fluctuations in the orientation of the eccentricity and the
magnetic field direction, there is a substantial $\left< |B_{x}'| \right>$ component.
However, the potential for the three-point correlator to measure the CME remains, as the
two components are still significantly different (the mean absolute value
$\left< |B_{y}'| \right> \approx 1.5 \times \left< |B_{x}'| \right>$).

\begin{figure*}[ht!]
\centering
\includegraphics[width=0.95\linewidth]{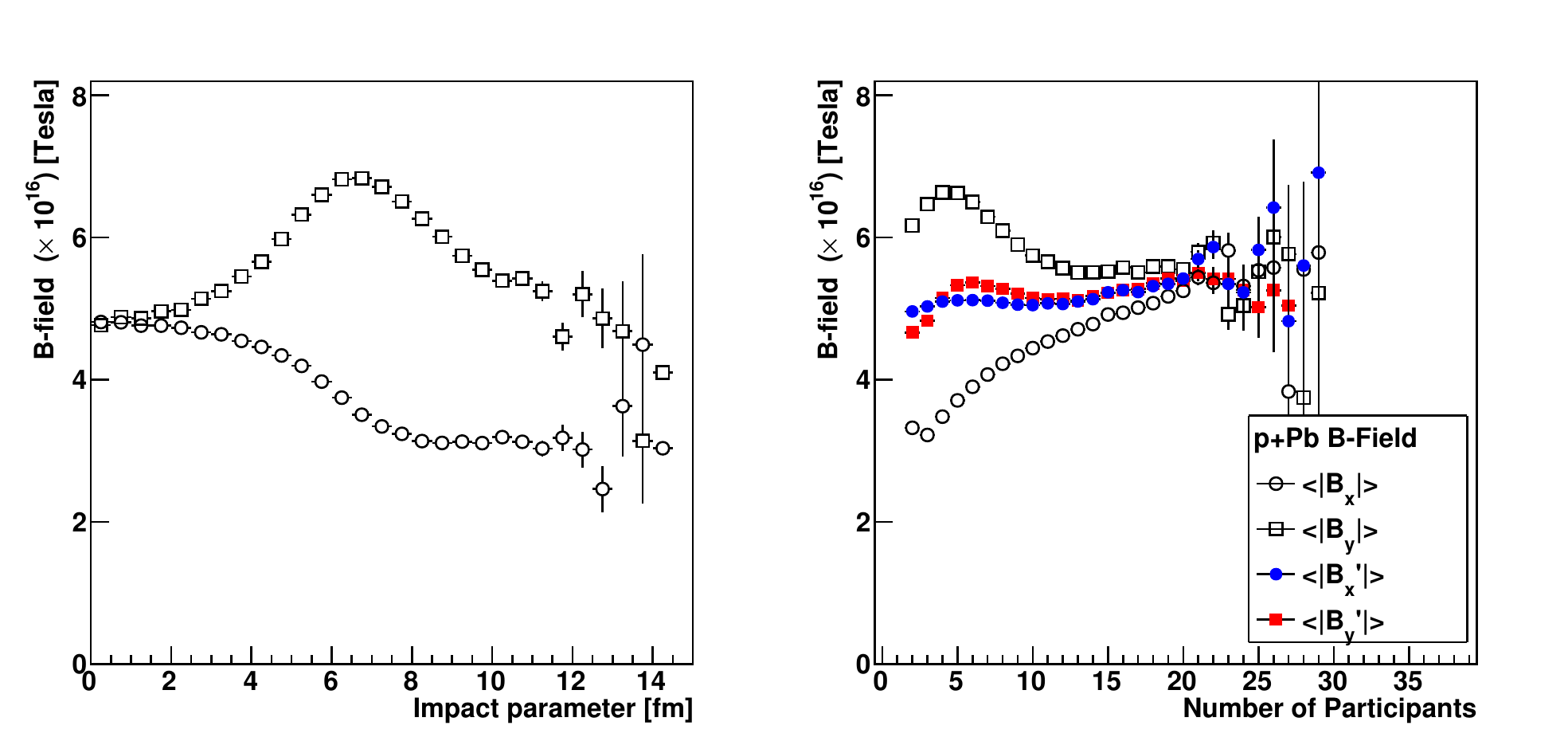}
\caption{Magnetic field components parallel ($|B_x|$, indicated by black open circles) and
  perpendicular ($|B_y|$, black open squares) to the impact parameter vector as a function of
  impact parameter (left panel) and $N_{part}$ (right panel) in p+Pb collisions.  The
  filled red squares on the right panel indicate the rotated components parallel to the eccentricity direction ($|B_x'|$) and
  the filled blue circles rotated component perpendicular to the eccentricity direction ($|B_y'|$).}
\label{fig:bfield_ppb}
\end{figure*}

Figure~\ref{fig:bfield_ppb} shows the same quantities but now for p+Pb collisions.  Two clear
conclusions can be reached.  First, the magnetic field mean absolute values are not small.
In fact, the magnetic field magnitudes in the
rotated frame are comparable to the Pb+Pb
$x'$ component and only about 50\% smaller than the $y'$ component.  The average impact parameter
even in the large number of participating nucleon events is of order 1.5-1.7 fm (still
non-zero) and the fluctuations in the nearest protons to the participant center-of-mass
generate significant fields.

More importantly, the second conclusion is that for p+Pb collisions, the
magnetic field direction and the eccentricity orientation are uncorrelated so that one
finds that $\left< |B_{y}'| \right> = \left< |B_{x}'| \right>$.  This confirms the second
postulate in the CMS paper, and gives the strong conclusion that the CME must not be
the effect being observed in the three-point correlator in p+Pb interactions.
Given the similarity of the p+Pb and Pb+Pb experimental results, it is also unlikely
that the Pb+Pb are the result of the CME.

\begin{figure*}[ht!]
\centering
\includegraphics[width=0.95\linewidth]{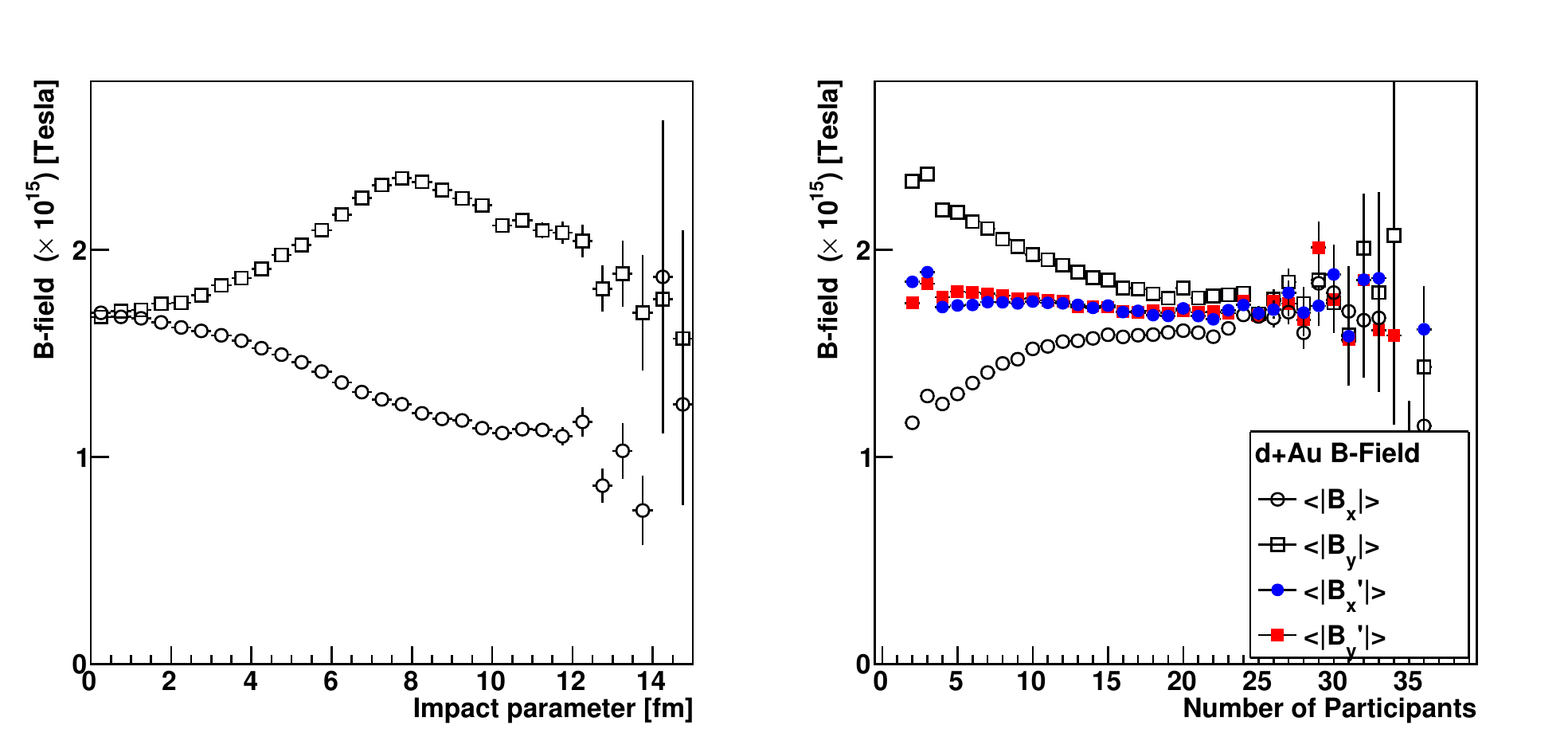}
\caption{Magnetic field components parallel ($|B_x|$, indicated by black open circles) and
  perpendicular ($|B_y|$, black open squares) to the impact parameter vector as a function of
  impact parameter (left panel) and $N_{part}$ (right panel) in d+Au collisions.  The
  filled red squares on the right panel indicate the rotated components parallel to the eccentricity direction ($|B_x'|$) and
  the filled blue circles rotated component perpendicular to the eccentricity direction ($|B_y'|$).}
\label{fig:bfield_dau}
\end{figure*}

Figure~\ref{fig:bfield_dau} shows the same quantities but now for d+Au collisions
at $\sqrt{s_{NN}}$ = 200 GeV; note that
the arbitrary scale is different from that of Figures~\ref{fig:bfield_pbpb}~and~\ref{fig:bfield_ppb}.
The exact same conclusions about the magnetic field in p+Pb can be drawn about that in d+Au.
The study of this correlator in small systems at RHIC energies should be elucidating.
We note that in 2016 d+Au data was taken at collision energies 200, 62.4, 39, and 19.6 GeV.


\begin{figure*}[ht!]
\centering
\includegraphics[width=0.49\linewidth]{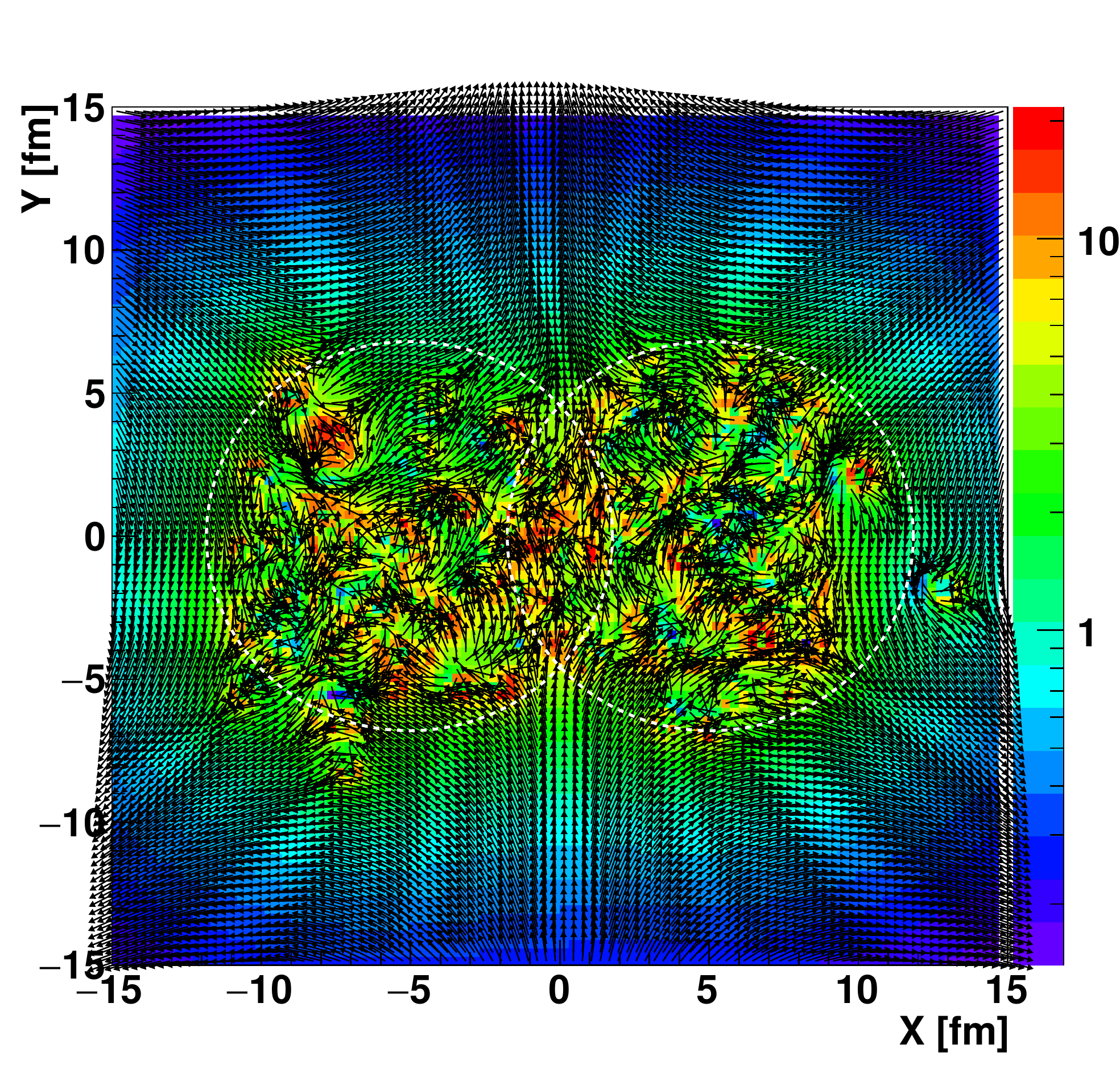}
\includegraphics[width=0.49\linewidth]{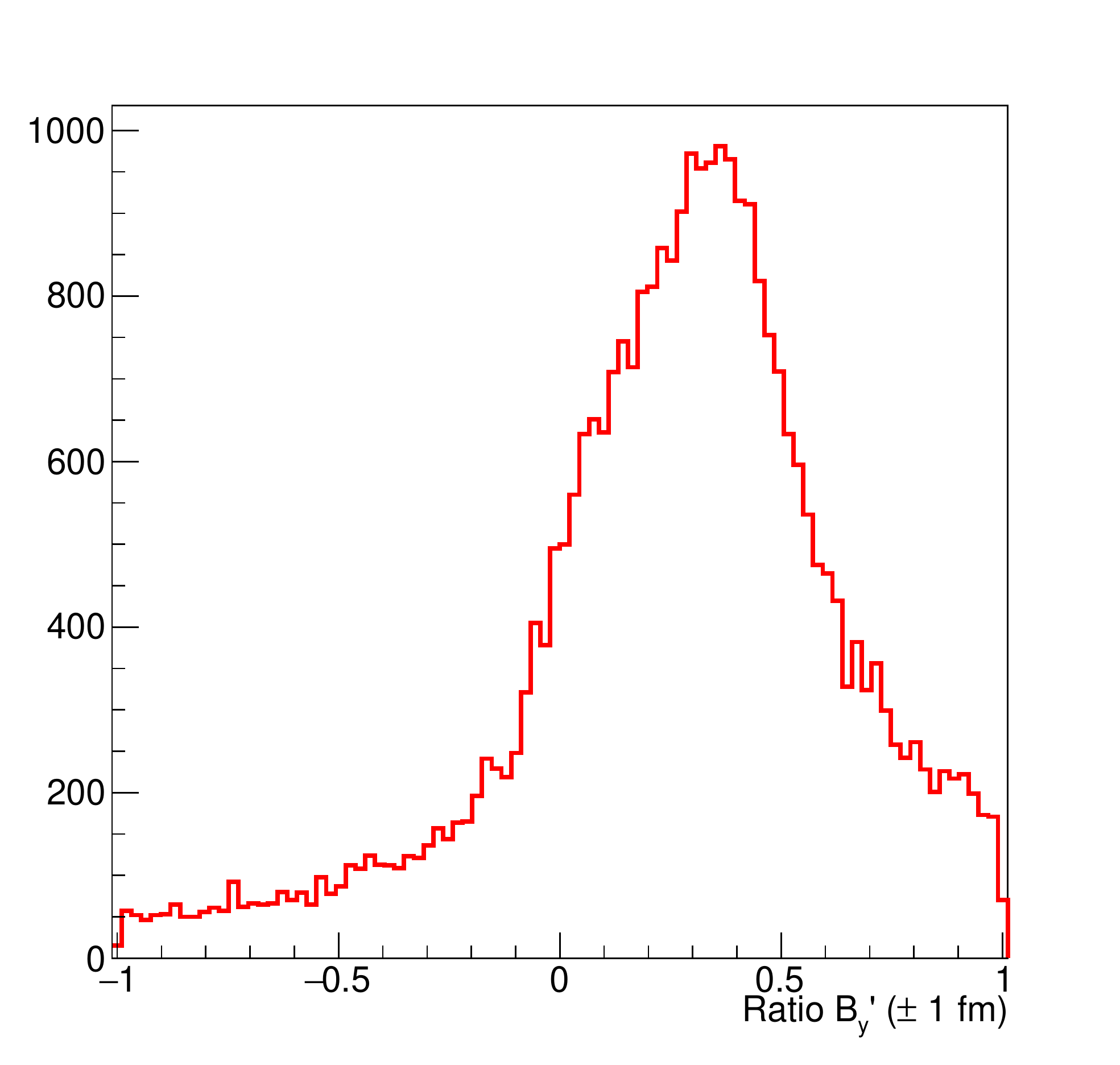}
\caption{(Left panel) Magnetic field map in a peripheral Pb+Pb collisions, where the
  color indicates the field strength and the arrow indicates the direction.
  The preponderence of more red colors in the overlap zone indicates a stronger magnetic
  field in that area.
  (Right panel) Ratio of magnetic field component perpendicular to the eccentricity
  direction ($|B_y'|$) at one point and at another point 1~fm away along the same
  direction.  A positive value indicates the fields at the two points are aligned,
  a negative value indicates the fields are anti-aligned.}
\label{fig:bfield_homogeneity}
\end{figure*}

It is notable that even in Pb+Pb semi-central collisions, where the average magnetic field
and the flow vector retain a correlation, the magnetic field is highly inhomogeneous.
The left panel of Figure~\ref{fig:bfield_homogeneity} shows a single Pb+Pb collision,
where the color scale in log $z$ shows the magnitude of the magnetic field as a map in
transverse position space.  The arrows indicate the direction of the magnetic field in
each spatial cell.
The preponderance of red color in the overlap region and lighter colors away from there
is qualitatively consistent with general expectations.  On the other hand, the field directions
do have significant fluctuations away from the average in the overlap region, indicating a
large degree of inhomogeneity.
In order to quantify the inhomogeneity, we examine Pb+Pb collisions
with 75~$<N_{part}<$~100, averaging over an ensemble of events.  We compute the component of the magnetic
field along the flow vector ($|B_y'|$) at two positions: 1 fm above the center of mass and
1 fm below the center of mass.  We then compute the ratio of these two values with the
larger magnitude number always in the denominator.  A positive value indicates the components
along the flow vector are in the same direction; a negative value indicates the components
are in opposite directions.  This quantity is shown in the right panel of Figure~\ref{fig:bfield_homogeneity}.
Approximately 20\% of the time, the components are in the opposite direction whereas they are
in the same direction about 80\% of the time.  However, even when the components are in the same
direction, the value of the smaller of two is about half that of the larger.
Any calculation of the CME should take into consideration the inhomogeneities of the magnetic field,
which have roughly the same size scale as the topological domains.

\section{Summary and outlook}
The impact of the CMS result is clear.  As we have demonstrated in this paper,
the magnetic field direction in p+Pb
collisions is uncorrelated with the flow angle, and therefore the CME cannot contribute any signal to
the observed correlations.
Considering the remarkable similarity between the p+Pb and
Pb+Pb measurements, any CME contribution to the observed
heavy ion collisions at the LHC must be heavily subdominant.
While the case is likely quite different at RHIC, where the
magnetic field is weaker but much longer lived, the similarity between the
centrality dependence
LHC and RHIC
results discussed earlier remains an important open question.
Because the CME is subdominant to a mimic signal, a much higher burden of proof is required for
validation.
The theory behind the CME is
extremely strong.  The physics of the $U(1)_A$ anomaly in QCD is very well established and
very clear.  If we had an arbitrarily long-lived quark gluon plasma and could embed it in
an arbitrarily strong and long-lived magnetic field, there is essentially no question that
the CME would occur.  Indeed, there is an analogous effect in QED (to
wit, the axial anomaly in QED allows the neutral pion to decay into two photons), which
has not only been predicted but observed~\cite{Li:2014bha}.  Regrettably, the
quark gluon plasma we can create in the lab has relatively short life time, and the
magnetic field induced is the strongest magnetic field observed anywhere in the universe
but also extremely short lived.  However, while it is clear that the CME must be heavily
subdominant to the observed correlations, there is still the possibility that
sufficiently advanced detectors and techniques can observe it.
As Eugene Wigner once
said, the optimist regards the future as uncertain.

\begin{acknowledgments}
We acknowledge funding from the Division of Nuclear Physics of the
U.S. Department of Energy under Grant No. DE-FG02-00ER41152.
\end{acknowledgments}

\bibliography{refs.bib}

\end{document}